\documentclass[12pt]{article}
\usepackage[dvips]{epsfig}
\begin{document}
\begin{flushright}
OHSTPY-HEP-T-98-021 \\
hep-th/9809165
\end{flushright}
\vspace{20mm}
\begin{center}
{\LARGE On Exact Supersymmetry in DLCQ }
\\
\vspace{20mm}
{\bf F. Antonuccio, O. Lunin,  and  S. Pinsky \\}
\vspace{4mm}
Department of Physics,\\ The Ohio State University,\\ Columbus, OH 43210, USA\\
\vspace{4mm}
\end{center}
\vspace{10mm}

\begin{abstract}

In recent years a supersymmetric form of discrete light-cone quantization
(hereafter `SDLCQ') has emerged as a very powerful tool for solving
supersymmetric 
field theories. In this scheme, one 
calculates the light-cone supercharge with respect to a
discretized light-cone Fock basis, instead of working 
with the light-cone Hamiltonian. 
This procedure
has the advantage of preserving supersymmetry even in 
the discretized theory, and eliminates
the need for explicit renormalizations in 
$1+1$ dimensions.
In order to compare the usual DLCQ prescription with 
the supersymmetric prescription,  
we consider two dimensional  SU($N$) Yang-Mills theory
coupled to a massive adjoint Majorana fermion,
which is known to be supersymmetric at a particular value of
the fermion mass. After studying how singular-valued
amplitudes and intermediate zero momentum modes are regularized in
both schemes, we are able to establish a precise connection
between conventional DLCQ and its supersymmetric extension, SDLCQ.
In particular, we derive the explicit form of the (irrelevant) 
interaction that renders the DLCQ formulation
of the theory exactly supersymmetric for any light-cone compactification.
We check our analytical results via a numerical procedure, 
and discuss the relevance of this
interaction when supersymmetry is explicitly broken.
\end{abstract}
\newpage

\baselineskip .25in

\section{Introduction}
The properties of 
supersymmetric gauge theories
are of particular interest nowadays
because of intriguing connections that are believed to exist
between large $N$ super Yang-Mills theories and
string/M theory \cite{witt96,bfss97,mald97,tay98}. 
A string/M theoretic interpretation of a class of super Yang-Mills
theories at finite $N$ was also provided by Susskind \cite{suss97},
and it was in this context that DLCQ emerged
as an interesting conceptual tool in the non-perturbative formulation
of M theory.


Remarkably, DLCQ turns out to be very
useful in practical bound state calculations \cite{pb85}, and 
this fact has been readily
exploited in the context of many studies 
of two dimensional field theories
(see \cite{bpp98} for a review). 
In more recent times, a DLCQ prescription preserving 
exact supersymmetry for any
discretization of the total 
light-cone momentum (originally proposed in \cite{mss95}) has been 
employed in
a study of a large class of two dimensional
supersymmetric matrix models (\cite{hak95,alp98,alp98b,alppt98,appt98}). 

Despite these developments, a detailed knowledge of 
DLCQ in the context of supersymmetric theories is still lacking,
although steps in that direction have been 
taken \cite{hellerman97,burk98}.
In the present paper, we delve into this issue further
by investigating the relation between 
conventional DLCQ,  and the supersymmetric 
form of DLCQ (hereafter `SDLCQ') proposed in \cite{mss95}. 
For this purpose, we employ both DLCQ prescriptions in a study of 
an SU($N$) gauge theory coupled to a single Majorana adjoint
fermion, which is known to be supersymmetric
for a particular value of the fermion mass\footnote{
A numerical study of this model using conventional DLCQ  
was initially carried out in \cite{dk93}. A subsequent 
study of the model at the supersymmetric
point can be found in \cite{bdk93}.}
\cite{kut93,bok94}.

The content of this paper may be summarized as follows;
in Section \ref{zero}, we establish a precise connection between
conventional DLCQ and SDLCQ for an SU($N$) 
gauge theory coupled to an adjoint Majorana fermion.
It turns out that the relationship between the two prescriptions
hinges on the regularization of a singular amplitude
governing a two-body $\rightarrow$ two-body interaction with zero
momentum exchange. The difference in the two prescriptions
may be encoded as an operator, and we present its explicit
form.  In Section \ref{numerical}, we test our
analytical results with a numerical analysis, and
study the behavior of this operator at the supersymmetric point,
and when supersymmetry is broken explicitly.
Different boundary conditions for the fermions are also 
considered. We conclude in Section \ref{discussion} with a summary of 
our results, and speculate on the enhanced significance 
of DLCQ when supersymmetry is present.

\section{\bf Zero Modes and Supersymmetric Regularization.}
\renewcommand{\theequation}{1.\arabic{equation}} \setcounter{equation}{0}
\label{zero}

We consider the $1+1$ dimensional SU($N$) gauge theory
coupled to an adjoint
Majorana fermion.
The light-cone quantization of this model in the light-cone
gauge and large $N$ limit has been dealt with explicitly 
before \cite{dk93,bdk93},
and we choose to adopt the same notation here\footnote{
For a treatment of this model at finite $N$, the
reader is referred to \cite{ap98}.}.
The expressions for the light-cone momentum $P^+$
and light-cone Hamiltonian $P^-$ for this model are 
\begin{eqnarray}
\label{momenta}
P^+ & = & \int dx^-
\mbox{tr}({\rm i}\psi\partial_-\psi),\\ P^- & = & \int
dx^- \mbox{tr}\left(-\frac{{\rm i}m^2}{2}\psi\frac{1}{\partial_-}\psi-
\frac{g^2}{2}J^+\frac{1}{\partial_-^2}J^+\right).
\end{eqnarray}
Here
$J^+_{ij}=2\psi_{ik}\psi_{kj}$ is the longitudinal current. 
It is well
known that at a special value of fermionic mass (namely
$m^2=g^2N/\pi$)
this system is supersymmetric \cite{kut93}. 
This special value of the fermion mass will be denoted by $m_{SUSY}$.
At this supersymmetric point, the supercharge is given by 
\begin{equation}
\label{sucharge}
Q^-=2^{1/4}\int dx^-\mbox{tr}(2\psi\psi\frac{1}{\partial_-}\psi)
\end{equation}
which satisfies the supersymmetry relation $\{Q^-,Q^-\}=2\sqrt{2}P^-$. 
This may
be checked explicitly by using the anticommutator at equal $x^+$:
\begin{equation}
\{\psi_{ij}(x^-),\psi_{kl}(y^-)\}=
\frac{1}{2}\delta_{il}\delta_{jk}\delta(x^- -y^-).
\end{equation}
In the DLCQ formulation, the theory is regularized  by
a light-like compactification, 
and either periodic or antiperiodic boundary conditions
may be imposed for fermions.
If $P^+$ denotes the total light-cone momentum, light-like 
compactification is equivalent to restricting 
the light-cone momentum of partons to be non-negative 
integer multiples of $P^+/K$, where $K$ is some positive
integer that is sent to infinity in the decompactified limit\footnote{
$K$ is sometimes called the `harmonic resolution', or just `resolution'.}.
Anti-periodic boundary conditions will in general explicitly break the
supersymmetry in the discretized theory,
although supersymmetry will be restored in the
decompactification limit $K\rightarrow
\infty$ \cite{bdk93}.  
 If we wish to maintain supersymmetry 
at any finite $K$, we must at least impose
periodic boundary conditions for the fermions.
This, however, leads to the notorious ``zero-mode problem''\footnote{
 For anti-periodic boundary conditions,
the light-cone momentum of partons is restricted to 
{\em odd} integer multiples of 
$P^+/K$, and so there are no zero-momentum modes in such a formulation.}.
From a numerical perspective, omitting zero-momentum modes
in our analysis is absolutely necessary, since 
it guarantees a {\em finite} Fock basis for each finite resolution $K$.
The mass spectrum of the continuum theory may then be extracted 
after an appropriate extrapolation of masses obtained from
diagonalizing a sequence of finite mass matrices for $M^2=2P^+P^-$.
But are we really justified in omitting the zero-momentum modes?
To date, the general consensus is that omitting
zero momentum modes in a two dimensional interacting field
theory does not affect the spectrum of the decompactified
theory, where 
$K \rightarrow \infty$.\footnote{In the
continuum theory, the (arbitrarily small) region 
surrounding the zero mode $k^+=0$ becomes important, rather than the single 
mode $k^+=0$, which has measure zero.}
Actually, the numerical results of Section \ref{numerical} are 
consistent with this viewpoint.

However, the goal of this work is to understand
the structure of a supersymmetric theory at finite resolution. 
As we will see shortly, understanding why the DLCQ
and SDLCQ prescriptions differ involves studying certain 
intermediate zero-momentum processes. But first,
we need to be more precise about the form of the
light-cone operators of the theory.
If we expand the fermion field $\psi_{ij}$ in terms of its
Fourier components, we may express the uncompactified
light-cone supercharge
and Hamiltonian in a momentum space representation
involving fermion creation and annihilation operators:
(\cite{kut93,dk93,bdk93}):
\begin{eqnarray}
\label{qminus}
Q^-&=&\frac{i2^{-1/4}g}{\sqrt{\pi}}\int_0^\infty dk_1dk_2dk_3
\delta(k_1+k_2-k_3)
\left(\frac{1}{k_1}+\frac{1}{k_2}-\frac{1}{k_3}\right)\times\nonumber\\
&\times&\left(b^\dagger_{ik}(k_1)b^\dagger_{kj}(k_2)b_{ij}(k_3)+
b^\dagger_{ij}(k_3)b_{ik}(k_1)b_{kj}(k_2)\right),\nonumber\\
& &  \\
\label{pminus}
P^-&=&\frac{m^2}{2}\int_0^\infty\frac{dk}{k}b^\dagger_{ij}(k)b_{ij}(k)+
\frac{g^2N}{\pi}\int_0^\infty\frac{dk}{k}\int_0^k dp \frac{k}{(p-k)^2}
b^\dagger_{ij}(k)b_{ij}(k)+\nonumber\\ &+&\frac{g^2}{2\pi}\int_0^\infty
dk_1dk_2dk_3dk_4\left[ \frac{}{}\delta(k_1+k_2-k_3-k_4)
A(k)b^\dagger_{kj}(k_3)b^\dagger_{ji}(k_4)b_{kl}(k_1)b_{li}(k_2)+\right.
\nonumber\\
&+&\left.\delta(k_1+k_2+k_3-k_4)B(k)(
b^\dagger_{kj}(k_4)b_{kl}(k_1)b_{li}(k_2)b_{ij}(k_4)-
b^\dagger_{kj}(k_1)b^\dagger_{jl}(k_2)b^\dagger_{li}(k_3)b_{ki}(k_4))
\frac{}{}\right]\nonumber
\\
& & 
\end{eqnarray}
with
\begin{eqnarray}
A(k)=\frac{1}{(k_4-k_2)^2}-\frac{1}{(k_1+k_2)^2},\nonumber\\
B(k)=\frac{1}{(k_3+k_2)^2}-\frac{1}{(k_1+k_2)^2}.
\end{eqnarray}
As we mentioned earlier,
the continuum theory is supersymmetric 
for a special value of fermion mass. We
will therefore consider only the case $m=m_{SUSY}$.
In the DLCQ formulation, one simply  
restricts integration of the light-cone momenta $k_i$
in expression (\ref{pminus}) for $P^-$
above to be {\em positive} integer multiples of $P^+/K$.
i.e. one simply drops the zero-momentum mode.
The DLCQ mass spectrum is then obtained by diagonalizing
the mass operator $M^2=2P^+P^-$.
Similarly, in the SDLCQ formulation, the 
light-cone momenta $k_i$
in expression (\ref{qminus}) for $Q^-$
are restricted to {\em positive} integer multiples of $P^+/K$.
One then simply {\em defines} $P^-$ to be the square
of the supercharge: $2 \sqrt{2} P^- = \{Q^-,Q^- \}$.
The mass operator $M^2=2P^+P^-$ is then easily constructed
and diagonalized to obtain the
SDLCQ spectrum.

In general, the following observations are made; at finite
resolution, the DLCQ spectrum of a supersymmetric theory 
is not supersymmetric. However, supersymmetry is restored
after extrapolating to the continuum limit $K \rightarrow \infty$
(see \cite{bdk93}, for example). In contrast, for any
finite resolution, the SDLCQ spectrum is supersymmetric.
The DLCQ and SDLCQ spectra agree only in the decompactified 
limit $K \rightarrow \infty$. 

Not surprisingly, the difference in the
DLCQ and SDLCQ prescriptions at finite resolution
may be understood as a zero-mode contribution. 
What is surprising is that we can
encode the effect of these zero-mode contributions into
a simple well defined operator.
The main result of this paper will be to state 
the precise operator form of this contribution at finite $K$.

In order to motivate our argument, note that the 
anticommutator  for the 
supercharge $Q^-$ in the continuum theory involves
products of terms of the form
$b^\dagger(k)b^\dagger(0)b(k)$ and $b^\dagger(p)b(0)b(p)$, and these 
provide contributions to $P^-$ that may be expressed in terms 
of non-zero momentum modes.
The problem is
exacerbated by the fact that the coefficients of these terms behave
singularly. To
examine this more closely, we consider
the discretized theory
where the light-cone momenta $k_i$ in the expression
for $Q^-$ [eqn(\ref{qminus})] are restricted to positive integer
multiples of $P^+/K$. We also include the 
effects of zero-momentum modes by introducing
an `$\epsilon$ regulated zero mode', which are modes
with momentum $k_i = \epsilon$, where $\epsilon$ is
much less than $P^+/K$, and is sent to zero at the end of
the calculation. 

Then the anticommutator of $Q^-$
gives contributions
 of the following form,
\begin{eqnarray}
&&\left\{(\frac{1}{\epsilon}+\frac{\epsilon}{k(k+\epsilon)})
b^\dagger(k)b^\dagger(\epsilon)b(k+\epsilon),
(\frac{1}{\epsilon}+\frac{\epsilon}{p(p+\epsilon)})
b^\dagger(p+\epsilon)b(\epsilon)b(p)\right\}=\\
&&=b^\dagger(k)b(k+\epsilon)b^\dagger(p+\epsilon)b(p)
\left[\frac{1}{\epsilon^2}+(\frac{1}{p(p+\epsilon)}+\frac{1}{k(k+\epsilon)})
+\frac{\epsilon^2}{pk(p+\epsilon)(k+\epsilon)}\right],\nonumber
\end{eqnarray}
where any terms involving
an $\epsilon$ regularized zero mode on the right-hand-side are 
dropped\footnote{In the DLCQ formulation of $P^-$, we omit zero modes.}. 
We have suppressed all matrix indices in this expression. In the limit
$\epsilon\rightarrow 0$ the last term on the right-hand-side
in the brackets vanishes, 
while the first term
is the pure momentum--independent divergence
that was identified in an earlier study of this model
\cite{bdk93}, and is canceled if we adopt
a principal value prescription for singular amplitudes
in the definition of $P^-$. The second term however,
is clearly a finite contribution to $P^-$,
although it arises from the $\epsilon$ regulated zero modes
in $Q^-$, which are not present in the SDLCQ prescription
for defining $Q^-$. Consequently, in order to ensure the 
supersymmetry relation
$\{Q^-,Q^- \} = 2 \sqrt{2} P^-$
in the discretized formulation,
we must include an $\epsilon$ regularization of the
zero modes in the definition for $Q^-$,  and
then apply a principal value prescription in the presence
of any singular processes to eliminate $1/\epsilon$ divergences.

Stated slightly differently, we may decompose the supercharge into a part
without zero
modes $Q^-_{SDLCQ}$ (i.e. $k_i = nP^+/K, n=1,2,\dots$),
and terms with $\epsilon$ regularized zero modes,
$Q^-_\epsilon$. 
The anti-commutator $\{Q^-_{SDLCQ},Q^-_{\epsilon}\}$
contains only terms with $\epsilon$
regulated zero-modes. Since 
$Q^-=Q^-_{SDLCQ}+Q^-_{\epsilon}$ one finds 
\begin{equation}
\label{twoQ}
\{Q^-_{SDLCQ},Q^-_{SDLCQ}\}=
2\sqrt{2}P^-_{SDLCQ}= 2\sqrt{2}P^-_{DLCQ}-\{Q^-_{\epsilon},
Q^-_{\epsilon}\}_{PV},
\end{equation}
after dropping any $\epsilon$ regulated zero-mode terms 
in the calculated expression for $\{Q^-,Q^-\}$.
Note that the first equality above is just the definition
for the light-cone Hamiltonian $P^-$ in the SDLCQ prescription. 
The $PV$ abbreviation on the right hand side indicates a principal
value
regularization prescription,  which is tantamount to
dropping all $1/\epsilon$ terms as $\epsilon\rightarrow 0$. The
procedure is well known in the context of the present model 
\cite{bdk93}. It is clear that our definition for $P^-_{SDLCQ}$  
gives rise to the supersymmetry relation $[Q^-_{SDLCQ},P^-_{SDLCQ}]=0$,
which yields a supersymmetric spectrum for any finite resolution
$K$. Moreover, we know that
$P^-_{SDLCQ}$ and $P^-_{DLCQ}$ yield the same spectrum\footnote{This
fact is expected from physical grounds, but one can also check this 
numerically.} 
in the continuum limit $K\rightarrow \infty$,
so it remains to calculate the difference at finite resolution $K$.
 We will write this
difference in terms of their respective
mass operators: $M^2=2P^+P^-$. A straightforward
calculation of the
anticommutator on the right-hand-side of (\ref{twoQ}) leads to the result:
\begin{eqnarray}
\label{Mdiff}
M^2_{SDLCQ}-M^2_{DLCQ}&=&M^2_{\Delta}=-\frac{g^2NK}{\pi}\sum_n \frac{1}{n^2}
B^\dagger_{ij}(n)B_{ij}(n)\nonumber\\ && -\frac{g^2NK}{\pi}\sum_{mn}
(\frac{1}{m^2}+\frac{1}{n^2})\frac{1}{N}
B^\dagger_{kj}(m)B^\dagger_{ji}(n)B_{kl}(m)B_{li}(n).
\end{eqnarray}
We also write down
the expression for $M^2_{DLCQ}$ in the theory with periodic fermions:
\begin{eqnarray}
M^2_{DLCQ}&=&\frac{g^2NK}{\pi}\sum_n B^\dagger_{ij}(n)B_{ij}(n)(\frac{x}{n}+
\sum_m^{n-1}\frac{2}{(n-m)^2})+\nonumber\\ &+&\frac{g^2K}{\pi}\sum_{n_i}\
'\left\{
\delta_{n_1+n_2,n_3+n_4}
\left[\frac{1}{(n_2-n_4)^2}-\frac{1}{(n_1+n_2)^2}\right]
B^\dagger_{kj}(n_3)B^\dagger_{ji}(n_4)B_{kl}(n_1)B_{li}(n_2) \right.\nonumber\\
&+&\delta_{n_1+n_2+n_3,n_4}
\left[\frac{1}{(n_2+n_3)^2}-\frac{1}{(n_1+n_2)^2}\right]\\
&(&\left.B^\dagger_{kj}(n_4)B_{kl}(n_1)B_{li}(n_2)B_{ij}(n_3)-
B^\dagger_{kj}(n_1)B^\dagger_{jl}(n_2)B^\dagger_{li}(n_3)B_{ki}(n_4))
\right\}.\nonumber
\end{eqnarray}
In this expression the variable $x=\frac{\pi m^2}{g^2N}$ is a dimensionless 
mass parameter, and for 
the supersymmetric point we have $x=1$. The sums are performed over positive
integers, $0<n_i<K$, and we employ a principal value prescription 
in sums labeled as $\sum'$, which implies that terms of the form
$1/(k-k)^2$ are dropped. In the SDLCQ procedure we calculate $Q^-$ which is
non-singular
and requires no principal value prescription.

The term $M^2_\Delta$ appears to be non-trivial 
due to the presence of
$B^\dagger B^\dagger BB$ terms on the right hand side of 
(\ref{Mdiff}). 
However, the action of this term on any
SU($N$) Fock state turns out to be equivalent 
to the first term, although with opposite sign, and twice the
magnitude.  
Thus the 
action of the right hand side of
(\ref{Mdiff}) is equivalent to the single quadratic operator:
\begin{equation}
 M^2_{\Delta}=\frac{g^2NK}{\pi}\sum_n \frac{1}{n^2}
B^\dagger_{ij}(n)B_{ij}(n).
\label{olegterm}
\end{equation}
Fortunately, we are able to test this analytical result by performing
direct numerical simulations of this model using both prescriptions,
and comparing the differences observed with the above prediction.
Interestingly,
although this result was derived for large $N$, 
agreement turns out to be perfect for both finite and large $N$,
which was verified using the finite $N$ DLCQ 
algorithms developed in \cite{ap98}.
We discuss this further in the next section.

\section{\bf Numerical Results}
\renewcommand{\theequation}{1.\arabic{equation}} \setcounter{equation}{0}
\label{numerical}

First, let us review the numerical results for this model which were
discussed in {\cite{bdk93}.\footnote{ 
Results for finite $N$ can be found in ref\cite{ap98}.}
The authors imposed
anti-periodic boundary conditions for the fermion 
(which guarantees the absence of a zero-momentum mode) and
showed that at the supersymmetric point,
the extrapolated ($K=\infty$) mass squared $M^2$ 
for the lightest fermion and boson
bound states are equal, and approximately 25.9
in units $g^2 N/\pi$.
The convergence was relatively slow, and the mass squared obtained 
at the highest resolution ($K=25$) was still $15\%$ from the
final extrapolated value.  
It was also stated that periodic boundary conditions reproduce the 
same result, although
convergence is much slower. Our numerical calculations are
consistent with these observations.

We then repeated the calculation in SDLCQ. We find of course exact
fermion-boson mass degeneracies  at
every resolution, and by calculating masses up
to resolution $K=10$, we find an extrapolated mass squared
$M^2= 26.4$, in units $g^2 N/\pi$. The mass squared 
calculated at the highest resolution ($K=10$) was only
$7\%$ from the extrapolated value. Moreover, the convergence of $M^2$ 
appeared to be very close to a linear function of $1/K$,
and so the extrapolated value can be determined to high precision.
 
We then repeated the calculation
using the usual DLCQ prescription for $P^-$,
but with the additional term (\ref{olegterm}) that was calculated
in the previous section. As predicted, this `modified DLCQ', and
the SDLCQ results just mentioned are in perfect agreement. 

Since the SDLCQ prescription (or DLCQ with an additional term 
(\ref{olegterm})) yields the same spectrum in the 
decompactified limit $K \rightarrow \infty$ as the standard DLCQ
prescription for $P^-$ (i.e. DLCQ without additional term), we conclude
that the operator (\ref{olegterm}) is irrelevant in the continuum 
limit $K\rightarrow \infty$, since it gives a zero contribution.
Its only role is to guarantee supersymmetry in the compactified theory. 

Interestingly, this situation changes when we choose to break the 
supersymmetry explicitly by choosing a fermion mass $m \neq m_{SUSY}$
(or adding a mass term to the definition of $P^-$ in the SDLCQ
prescription).
Namely, for a non-supersymmetric value of the fermion mass,
the DLCQ spectrum with and without the additional term (\ref{olegterm})
appear to differ even in the 
decompactified limit $K\rightarrow \infty$. 
Interestingly, small perturbations in fermion mass away from
the supersymmetric point $m=m_{SUSY}$ were considered by Boorstein and 
Kutasov
\cite{bok94}, and
an explicit formula was derived that measured the 
splittings in the fermion and
boson masses in terms of the supersymmetry-breaking mass.
This result may be checked explicitly in both the DLCQ and
SDLCQ prescriptions. Surprisingly, the mass splittings 
observed in both prescriptions agree extremely well
with the Boorstein-Kutasov analysis, although the absolute values
for the masses in both schemes differ away from the supersymmetric point.

It is tempting to suggest that the discrepancy in continuum masses
away from the supersymmetric point 
reflects extremely slow convergence,
and would disappear if we were able to probe larger enough
values for $K$. Our numerical analyses so far cannot confirm such
a view at present.
A more dramatic interpretation is to propose that  
there is a possible phase
transition at the supersymmetric point \cite{bok94}, and 
we might be comparing two different phases in this theory.

\section{\bf Discussion}
\renewcommand{\theequation}{1.\arabic{equation}} \setcounter{equation}{0}
\label{discussion}

We have considered the connection between supersymmetric discrete light-cone
quantization (SDLCQ)  and the standard
prescription for DLCQ in numerical calculations. 
SDLCQ preserves supersymmetry even in the discretized theory,
and requires no explicit renormalization of interactions.
In DLCQ, one works with the light-cone Hamiltonian $P^-$,
which involves amplitudes that must be regulated
(typically by the principal value prescription)
in addition to the usual regularization 
imposed by neglecting $k^+=0$
modes.
In contrast to the SDLCQ scheme, DLCQ does not preserve supersymmetry
at finite resolutions, although supersymmetry is restored in
the decompactified limit $K \rightarrow \infty$.

In a simple example of a two dimensional
SU($N$) gauge theory coupled to an adjoint Majorana fermion,
we determined an  operator that connects DLCQ (regularized
using principal value) to 
SDLCQ. We were able to check our results numerically,
and obtained exact agreement for finite and large $N$. 
The derivation of this 
operator involved a careful
treatment of the zero modes that are normally omitted
in SDLCQ, and the principal value regularization
prescription in DLCQ. In particular, we demonstrated that
certain zero mode contributions may be succinctly encoded as
an operator involving non-zero momentum modes. 

Our numerical results indicated that the operator 
(\ref{olegterm})
which
restores supersymmetry when added to the usual DLCQ Hamiltonian
is irrelevant in the decompactified limit, since both DLCQ
and SDLCQ schemes agree in this limit.
However, for a non-supersymmetric choice of fermion mass,
this operator appears to become relevant even in the continuum
limit. In particular, the extrapolated masses at $K=\infty$ 
using both schemes differ, although
the mass splittings between bosons and fermions 
in each scheme were in agreement with the analysis of ref\cite{bok94}.

One suggestion is that the model exhibits very bad convergence
behavior, although a detailed analysis conducted so far reveals 
very little evidence for this.
It is tempting to speculate 
that the zero-mode degrees of freedom
that are irrelevant at the supersymmetric point become relevant
when supersymmetry is broken.\footnote{ 
Evidence that zero-mode degrees of freedom may cancel when supersymmetry 
is present was provided recently \cite{burk98}.}
Clearly, this 
transition reflects a dynamical property of the theory, since 
one can only check the relevance or irrelevance of an operator by 
explicitly solving for the bound states in the theory
for different fermion mass, as we have done 
here. It would be interesting to connect this observation with
a proposal that the theory undergoes a phase transition at the 
supersymmetric point. Whether we are observing different phases 
in a theory would have very interesting ramifications in understanding
the nature of supersymmetry -- and perhaps supersymmetry breaking -- 
in a non-perturbative setting.

\section{\bf Acknowledgment}
This work was supported by a grant from the US Department of Energy. Two 
of the 
authors (F.A. and S.P.) would like to thank the Aspen Center for Physics
for its hospitality during part of this work.

\end{document}